\pdfoutput=1
\documentclass{JINST}
\usepackage[utf8]{inputenc}
\usepackage{amsmath}
\usepackage{graphicx}
\usepackage{url}
\usepackage[mathscr]{eucal}

\title{Wavelength-Shifting Performance of Polyethylene Naphthalate Films in a Liquid Argon Environment}
\author{Y. Abraham$^a$, J. Asaadi$^a$, V. Basque$^f$, W. Castiglioni$^b$, R. Dorrill$^b$, M. Febbraro$^d$, B. Hackett$^{d,e}$, J. Kelsey$^b$, B. R. Littlejohn$^b$, I. Parmaksiz$^a$, M. Rooks$^a$, A. M. Szelc$^c$ \\ \llap{$^a$} University of Texas at Arlington, \\ 701 S Nedderman Dr, Arlington, TX 76019, United States of America \\
\llap{$^b$} Illinois Institute of Technology, \\ 10 West 35th Street Chicago, Il 60616, United States of America \\
\llap{$^c$} University of Edinburgh, \\ Peter Guthrie Tait Road, Edinburgh, EH9 3FD, United Kingdom \\
\llap{$^d$} Oak Ridge National Laboratory, \\ 1 Bethel Valley Rd, Oak Ridge, TN 37830, United States of America \\
\llap{$^e$} University of Tennessee, \\ Knoxville, TN 37996, United States of America \\
\llap{$^f$} University of Manchester, \\Oxford Rd, Manchester M13 9PL, United Kingdom \\}

\abstract{
Liquid argon is commonly used as a detector medium for neutrino physics and dark matter experiments in part due to its copious scintillation light production in response to its excitation and ionization by charged particle interactions.  
As argon scintillation appears in the vacuum ultraviolet (VUV) regime and is difficult to detect, wavelength-shifting materials are typically used to convert VUV light to visible wavelengths more easily detectable by conventional means.  
In this work, we examine the wavelength-shifting and optical properties of poly(ethylene naphthalate) (PEN), a recently proposed alternative to tetraphenyl butadiene (TPB), the most widely-used wavelength-shifter in argon-based experiments.  
In a custom cryostat system with well-demonstrated geometric and response stability, we use 128~nm argon scintillation light to examine various PEN-including reflective samples' light-producing capabilities, and study the stability of PEN when immersed in liquid argon.  
The best-performing PEN-including test reflector was found to produce 34\% as much visible light as a TPB-including reference sample, with widely varying levels of light production between different PEN-including test reflectors.  
Plausible origins for these variations, including differences in optical properties and molecular orientation, are then identified using additional measurements.  
Unlike TPB-coated samples,  PEN-coated samples did not produce long-timescale light collection increases associated with solvation or suspension of wavelength-shifting material in bulk liquid argon.  
}

\keywords{Noble-liquid detectors; Photon detectors for UV}

\begin{document}

\maketitle
\section{Introduction} \label{sec:intro}

Large scale liquid argon (LAr) detectors play a central role in many aspects of high energy physics experiments.  
Liquid argon time projection chambers (LArTPCs) can be used to provide precise topological and calorimetric information about energetic charged products generated by interactions of particles of interest, be they dark matter or neutrinos.  
As charged particles traverse bulk argon material, they produce ionization electrons and scintillation photons.  
An external electric field directs ionization electrons towards a detector anode, where they can be collected directly on charge sensitive readout elements (such as wires or pads) or can be multiplied, converted into scintillation light, and subsequently detected in a gaseous argon phase.  
The combined measurement of the fast scintillation light, providing a reference interaction time, and the time of delayed ionization charge or scintillation light collection, allows for 3D reconstruction of the original ionization electron topology within the TPC.  
In addition to its utility in time projection chambers, LAr is an excellent scintillator, producing $\mathcal{O}(10,000)$ photons/MeV of deposited energy, making it an excellent candidate material for detectors seeking sensitivity to low energy recoils, such as dark matter experiments.  

A primary challenge in the use of LAr detectors is that its scintillation light is produced in the vacuum ultraviolet (VUV) range, with wavelengths well below 200~nm.  
While LAr is transparent to these wavelengths, most commercially available single photon sensitive optical detectors, such as silicon photomultipliers (SiPMs), multi-pixel photon counters (MPPCs), and photomultiplier tubes (PMTs), are largely insensitve to photons at these wavelengths. A common solution to this problem is the deployment of wavelength shifting (WLS) coatings, which can absorb the VUV photons and re-emit them at visible wavelengths where optical detectors are most efficient.  

One common WLS fluor is 1,1,4,4-tetraphenyl-1,3-butadiene (TPB), which has been deployed in a number of LAr experiments, such as ArDM~\cite{ArDM}, CENNS-10~\cite{cenns} DarkSide-50~\cite{darkside_first,darkside_light}, DEAP-3600~\cite{deap}, ICARUS~\cite{icarus_first}, LArIAT~\cite{lariat}, MicroBooNE~\cite{ub_det,ub_pmt_thesis}, WArP~\cite{warp}, and both protoDUNE single-phase~\cite{pd_sp_cdr,pd_sp_first} and dual-phase~\cite{pd_dp} detectors.   
TPB may or will be similarly used in future experiments that are either proposed, planned or currently under construction, such as the SBND and ICARUS detectors in the Fermilab SBN program~\cite{SBND1,ICARUS_deploy}, future argon-based dark matter experiments DarkSide-20k~\cite{darkside-20k} and Argo~\cite{GARDMC}, and DUNE~\cite{DUNETechDesign,DUNE_far}.  
TPB can be evaporatively coated or dissolved into a polymeric solution and applied to various surfaces, albeit at a degradation of WLS conversion.  
TPB has been studied to establish its absolute level of WLS efficiency at 128~nm \cite{Benson:2017vbw,Gehman:2011xm,Francini:2013lua}, its rate of and mechanisms for photo-degradation \cite{Chiu:2012ju_TPBDeg,acciarri2013aging}, and its emission time profile \cite{Segreto_TPB}.  
Recent work seeking to understand TPB's long-term stability in LXe~\cite{acciarri2013aging} and LAr~\cite{Asaadi:2018ixs} has also provided some initial evidence that TPB may not remain affixed to surfaces it is coated onto, which could cause unexpected WLS behavior in bulk argon as well as previously-observed degradation of LArTPC light collection over time~\cite{ub_degrade}.  
The complications involved in applying TPB to surfaces, as well as TPB's unclear long-term stability, have given rise to searches for alternative WLS materials.  

One candidate material which has recently gained attention is poly(ethylene naphthalate) (PEN). PEN is a semicrystalline thermoplastic polyester which can be extruded into plastic sheets, is relatively inexpensive, and has a wide industrial base of applications.  Intrinsically, PEN fluoresces in the blue region with a peak emission at 425 nm.  The emission is consists of two components, at 380  and 450 nm which correspond to the monomer and dimer respectably~\cite{excimer_DMN}. A series of recent studies of PEN~\cite{Molokanova} and its application to LAr experiments specifically~\cite{Kuzniak2019} has generated interest in the efficacy of PEN as an alternative to TPB~\cite{SzelcReview,LEGEND}. In this work we explore the relative WLS and optical performance of various samples of PEN-including reflectors, and compare these to the performance of a TPB-including reference sample in a LAr cryogenic environment, using 128 nm light from scintillation of the LAr itself. Additionally, we test the stability of both the TPB and PEN samples to see if there is any evidence of solvation or suspension in the bulk through seventy hour continuous runs of the LAr system.

This paper is organized as follows.  
Section \ref{sec:samples} describes the samples and their preparation as used in this study, while   Section \ref{sec:exp} describes the experimental setup, including its illumination sources, light collection devices, and readout electronics. 
Section \ref{sec:ProceduresSyst} defines the study's data-taking procedures and demonstrates the associated systematic errors.  
Section \ref{sec:results} provides the results of measurements made on various samples, and discusses implications and follow-up studies performed to more fully understand the acquired results.  
Finally, Section \ref{sec:discussion} presents a brief summary of the study's findings.  

\section{Sample Production and Characterization} \label{sec:samples}

\begin{figure}[htb]
    \centering
\includegraphics[trim = 0.0cm 0.0cm 0.0cm 0.0cm, clip=true, width=0.24\textwidth]{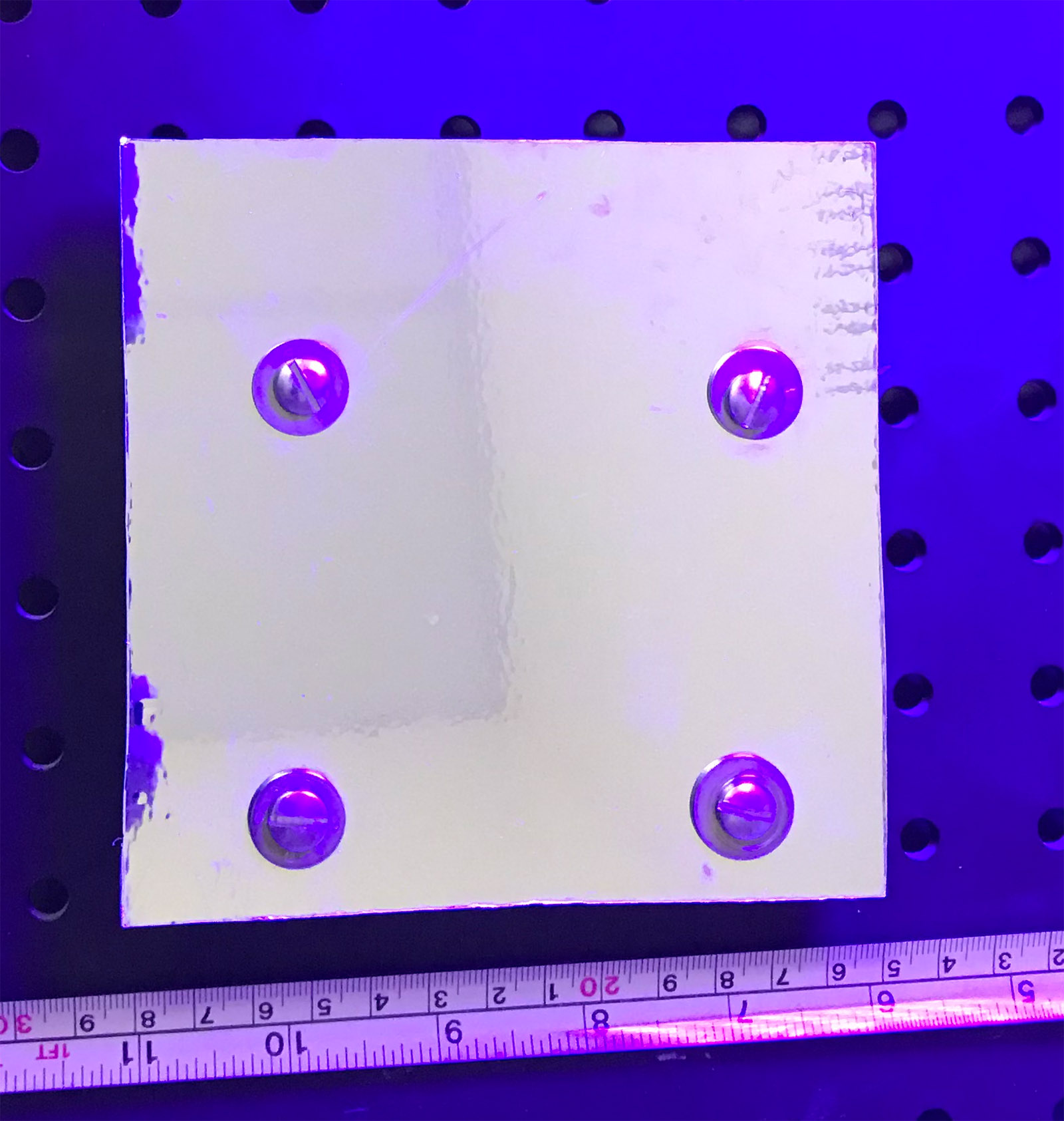} 
\includegraphics[trim = 0.0cm 0.0cm 0.0cm 0.0cm, clip=true, width=0.24\textwidth]{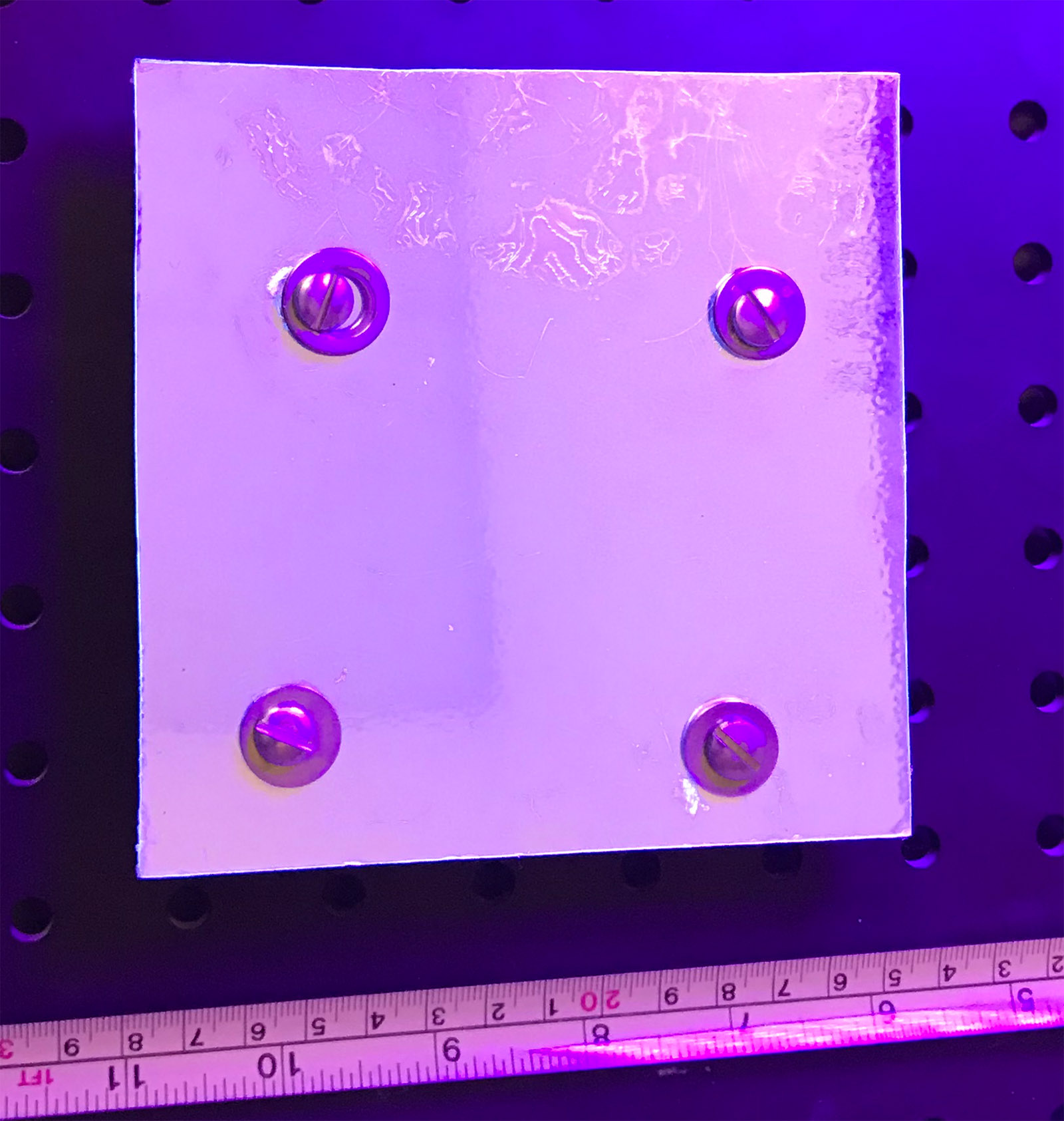} 
\includegraphics[trim = 0.0cm 0.0cm 0.0cm 0.0cm, clip=true, width=0.24\textwidth]{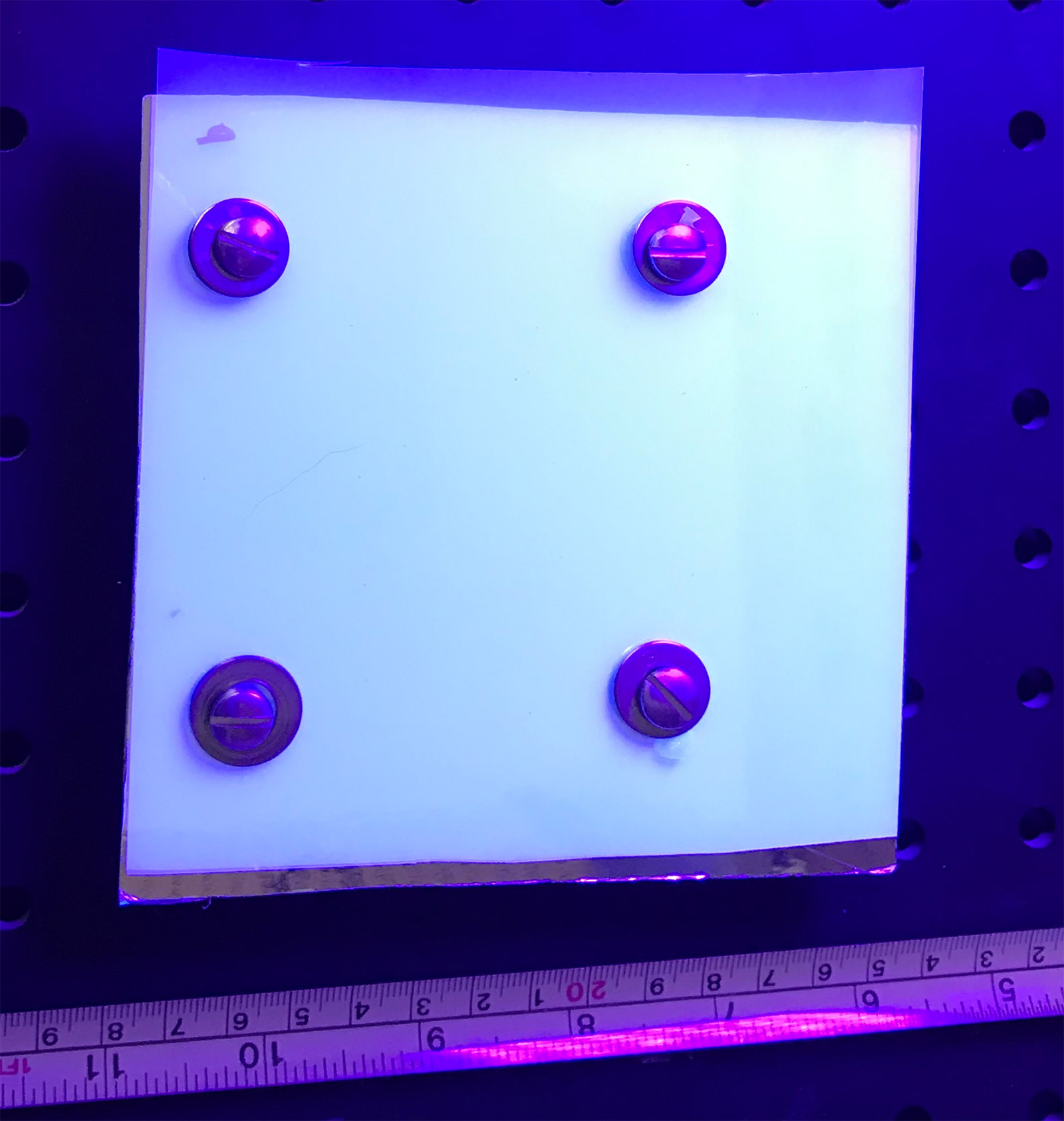} 
\includegraphics[trim = 0.0cm 0.0cm 0.0cm 0.0cm, clip=true, 
width=0.24\textwidth]{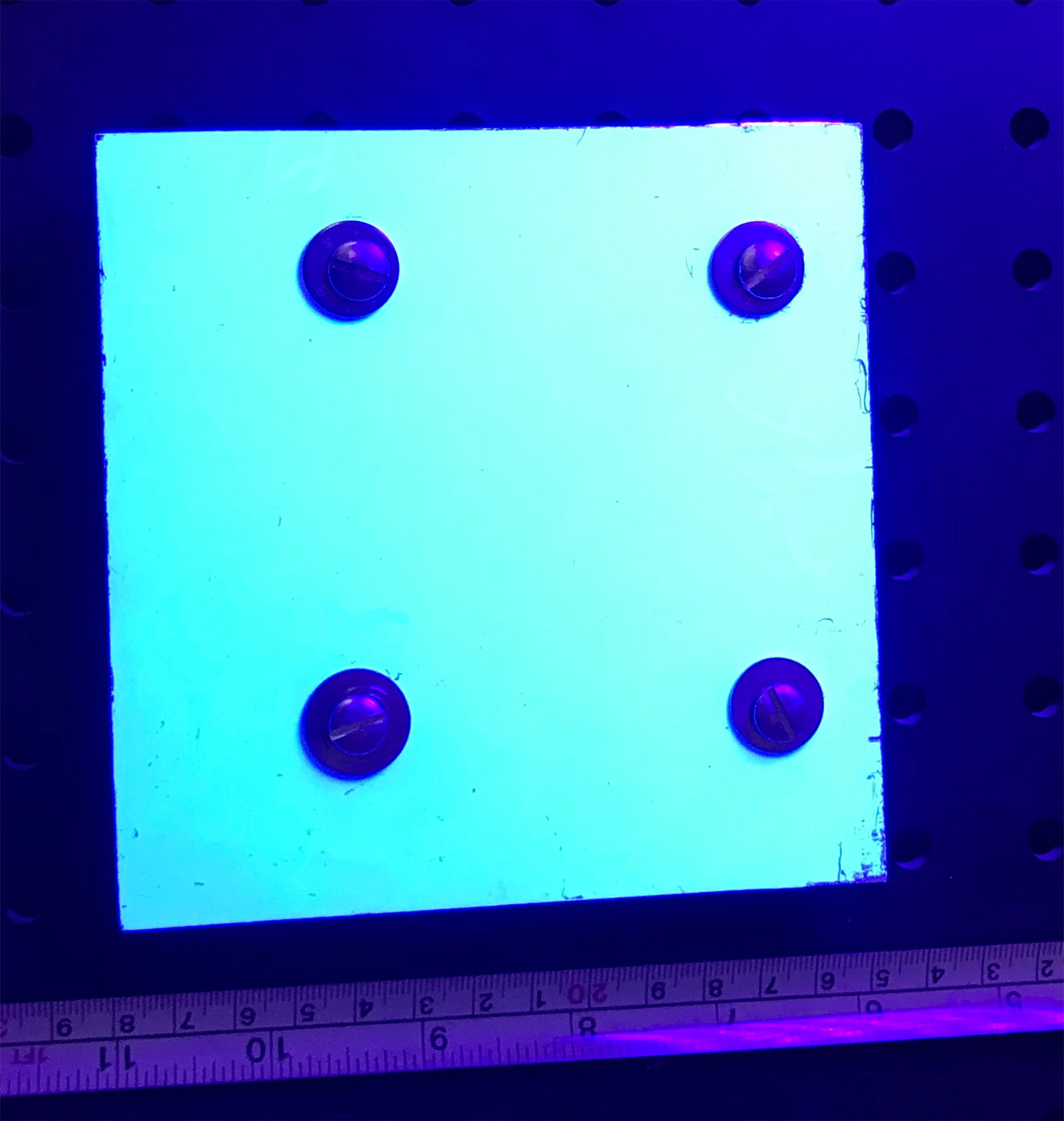} 
    \caption{Visual appearance under a UV lamp of the bare reflector (left) two PEN-containing samples (PEN01, left-center; PEN04, right-center) and one TPB-containing test sample (right).  While it is difficult to capture sample optical behavior in photographs, some differences are visible.  The clarity (haziness) of PEN material in PEN01 (PEN04) is visible by the similarity (difference) in reflected images between the left and center-left (center-right) images.  In addition, the lack of direct adhesion of PEN to reflector in sample PEN04 is clearly visible.  Finally, the diffuse character of the TPB coating is conveyed by the lack of a clear reflected image in the right-most image.  Detailed descriptions of these samples are provided in the section's text.}
    \label{fig:PicPen}
\end{figure}

A series of four PEN-including test samples and two PEN-excluding reference samples were produced and examined in this study.  
To achieve more consistent optics and subsequent light collection efficiency between measurements, all samples were produced as rigid, flat squares of 12.7$\times$12.7~cm$^2$ surface area.  
Wavelength-shifting candidate samples are also situated in front of a highly specularly reflecting surface to increase light collection, and to mimic the design and optics of WLS reflector panels appearing in  current and future LArTPCs, such as the SBND experiment at Fermilab~\cite{SBND1}.  
A backbone of glossy-surfaced, 0.8~mm thick FR4 fiberglass serves as a rigid base for each sample.  
DF2000MA, a 65~$\mu$m adhesive-backed specularly reflecting multilayer film produced by 3M~\cite{ESR} (also known by the trade name Enhanced Specular Reflector (ESR), and referred to as ESR throughout the article), is then pressure-laminated to this base using a hand roller.  
It should be noted that this reflective film contains many thin layers of PEN~\cite{ESR_science}, and has demonstrated a non-negligible level of capability as a VUV wavelength-shifter in previous studies~\cite{TPB_Gerda}.  
Some PEN samples are then pressure-laminated to this reflective layer using General Formulations CON106 optically clear double-sided adhesive in a manner similar to that reported in Ref.~\cite{optical_grid} for the optical grid of the PROSPECT liquid scintillator reactor neutrino detector.  
This acrylic-based adhesive binding ensures a fixed and consistent optical connection between WLS candidate and ESR reflective surface.  
Images of two PEN-including and one TPB-including sample are presented in Figure~\ref{fig:PicPen}.  

Two PEN-excluding reflector reference samples were produced to enable better quantification of PEN's performance as a VUV WLS material.  
The first consisted of a bare ESR reflector film with no test material attached to its front face.  
The second was evaporatively coated with TPB to a thickness of 0.3~mg/cm$^2$ in a large-volume evaporation chamber located at the University of Manchester, using procedures identical to those used to produce the TPB-coated samples described in Ref.~\cite{Asaadi:2018ixs}.  
The deposition method and thickness was similar to the ones used in the WArP 100l detector~\cite{WArP100}, as well as in the LArIAT~\cite{lariat} and SBND~\cite{SBND1} detectors. 
The product's performance is expected to be similar to coatings of higher densities~\cite{TPB_thick}, whereas they are mechanically somewhat more stable.  

All four tested PEN sample films, referred to as PEN01, PEN02, PEN03, and PEN04, were sourced from different vendors.  
The attributes of the different samples are summarized in Table~\ref{tab:SamplesList} and also are described below.  
Of the four sample films, PEN01, PEN03, and PEN04 were commonly manufactured by Dupoint Teijin and distributed under the Teonex trade name~\cite{TeonexPEN}, with some difference in associated product numbers, indicating some expected level of difference in physical properties.  
These three samples are all 0.125~mm thick and exhibit a smooth surface quality.  
Sample PEN03 was ordered from Millipore Sigma USA in August of 2018, and was supplied to them from a different distributor, Goodfellow USA; this sample is associated with the Teonex Q53 product number.  
PEN04 was ordered directly from Goodfellow in August of 2020 after direct shipping became available in the United States; this sample is in principle the same Q53 variety as that previously ordered from Millipore Sigma.  
Teonex Q53 is a biaxially oriented film that is visibly hazy in its optical appearance, as shown in Figure~\ref{fig:PicPen}.
In contrast, sample PEN01, procured from Piedmont Plastics in the United States, is associated with the Teonex Q65FA product number; it is described as low-crystallinity, and exhibits high optical transparency, as also shown in Figure~\ref{fig:PicPen}.  
As the impact of UV light exposure and extreme temperatures on the VUV wavelength shifting capabilities of PEN have not been quantified in the literature to our knowledge, samples were stored in light-tight containers at room temperature following purchase and fabrication. Despite detailed discussions with vendors, the age and environmental storage conditions of PEN films prior to purchase could not be ascertained.  

\begin{table}[]
    \centering
    \begin{tabular}{|c|c|c|c|c|}
    \multicolumn{5}{c}{Samples Tested}\\
    \hline
       \textbf{Data Sample}  & \textbf{Origin} & \textbf{Trade Name} & \textbf{Thickness} & \textbf{Properties} \\
       \hline
       \hline
        PEN01 & Piedmont & Teonex Q65FA  & 0.125m & Ultra-clear\\
        \hline
        PEN02 & ORNL & Teonex TN-8065S & 1.5mm  & Low crystallinity, clear\\ 
        \hline
        PEN03 & Millipore Sigma & Teonex Q53 & 0.125mm & Biaxially oriented, hazy\\
        \hline
        PEN04 & Goodfellow & Teonex Q53  & 0.125mm & Biaxially oriented, hazy\\ 
        \hline
        TPB & Manchester Univ. & - & 0.003mm & Evaporatively deposited \\
        \hline
        Bare & - &  - & - & No WLS layer applied \\
        \hline
    \end{tabular}
    \caption{The origins, names, thicknesses and properties of tested samples.  For all samples, provided origins, trade names, thicknesses, and properties refer to the wavelength-shifting material present on that sample.}
    \label{tab:SamplesList}
\end{table}

\begin{figure}[htb]
    \centering
    \includegraphics[trim = 0.0cm 0.0cm 0.0cm 0.0cm, clip=true, width=0.48\textwidth]{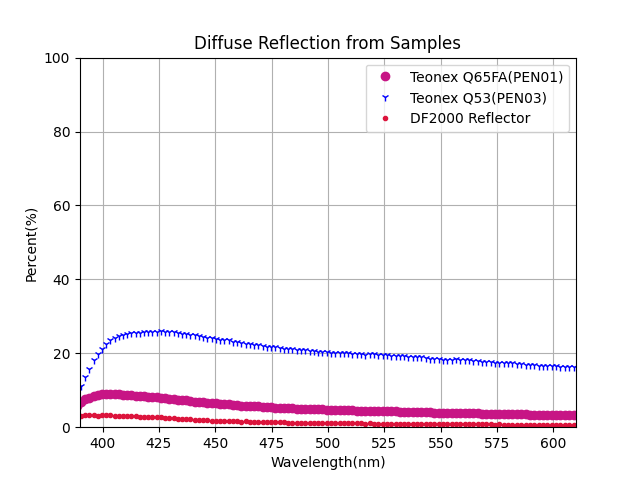} 
    \includegraphics[trim = 0.0cm 0.0cm 0.0cm 0.0cm, clip=true, width=0.48\textwidth]{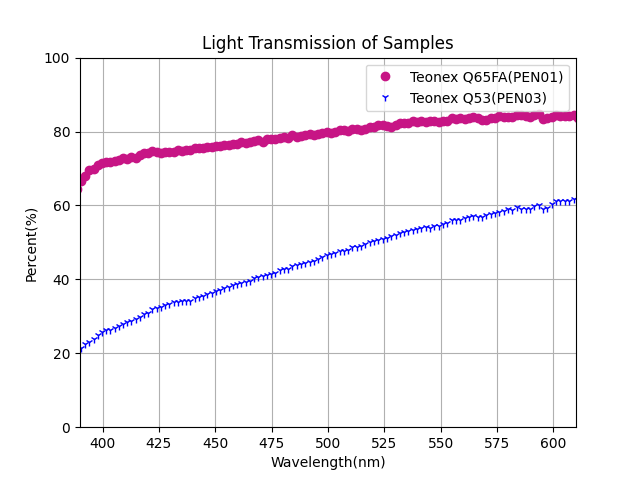} 
    \caption{Diffuse reflectance (left) and transmittance (right) measurements of two PEN samples and one ESR reflector sample as measured with an Ocean Optics STS-VIS UV-Vis spectrometer with fiber reflectance probe (left) and transmittance sample holder (right) attachments.}
    \label{fig:VisPen}
\end{figure}

To quantify the difference in optical performance between Teonex film samples, samples were examined using UV-Vis spectrometry, with measured transmittance and reflectance pictured in Figure~\ref{fig:VisPen}.  
Diffuse reflectances of bare PEN material from PEN01 and PEN03 and bare ESR material were measured using an Ocean Optics STS-VIS UV-Vis spectrometer~\cite{STS} with a fiber reflectance probe attachment.  
All samples were measured with a black cloth backing to absorb transmitted light, and then compared to a Spectralon 100\% diffusely reflecting standard.  
Few-percent measurement uncertainties are associated with the setup.  
The PEN-free reflector sample exhibits 1-2\% diffuse reflectance in the visible range, in good agreement with previous measurements on the same setup~\cite{optical_grid}.  
It can be observed that the Teonex Q65FA material present in the PEN01 sample exhibits optical quality similar to that exhibited by the ESR: both samples have few percent diffuse reflectance contributions. 
In contrast, the biaxially oriented Teonex Q53 PEN material present in PEN03 and PEN04 exhibits diffuse reflectances of order 10-25\% depending on the visible wavelength of interest.  

Transmittance of the individual Teonex films were also investigated with the same Ocean Optics STS-VIS system connected to a transmittance fiber attachment and a sample and fiber holder.  
Fractional visible transmittance results for Teonex Q53 and Q65FA are also shown in Figure~\ref{fig:VisPen}.  
The 0.125~mm thick Teonex Q65FA sample exhibits $>$70\% transmittance above 400~nm, with $\sim$89\% representing the maximum possible transmittance over two air ($n$=1 index of refraction) to PEN ($n\sim$1.6) interfaces.  
In contrast, Q53 Teonex exhibits only 25-60\% transmittance over the same wavelength range.  
This low transmittance relative to Q65FA can only be partially explained by its additional 10-20\% surface reflection and bulk scattering.  
As an example, at 400~nm, $\sim$20-25\% measured bulk diffuse reflection for Q53 cannot fully account for the $\sim$40\% lower transmittance relative to Q65FA.  
Thus, it seems likely that Teonex Q53 PEN exhibits both additional bulk scattering and additional bulk attenuation of visible light with respect to Teonex Q65FA.  

The final PEN-including sample, PEN02, used PEN supplied by Oak Ridge National Laboratory, which is described in detail in another publication~\cite{ORNL_PEN_2019}.  
The material is one of many samples produced as part of activities geared towards the development of scintillating radio-pure structural materials for the LEGEND neutrinoless double beta decay experiment~\cite{majoranaExp}.  
The PEN02 sample material was machined from a circular 1.5 mm thick PEN plate formed by injection compression-molding.  
The material used was sourced from Teonex TN-8065S following a low-background preparation and fabrication process.  
The result of this production processes was a PEN product with an amorphous (low-crystallinity) polymer structure, polished surface quality, and excellent optical clarity comparable to Teonex Q65FA.  
No attempt was made to modify the thickness of the original PEN product to more closely match that of the Teonex film sample materials.

\section{Experimental Setup} \label{sec:exp}

\begin{figure}[htb]
    \centering
\includegraphics[trim = 0.0cm 0.0cm 0.0cm 0.0cm, clip=true, width=0.465\textwidth]{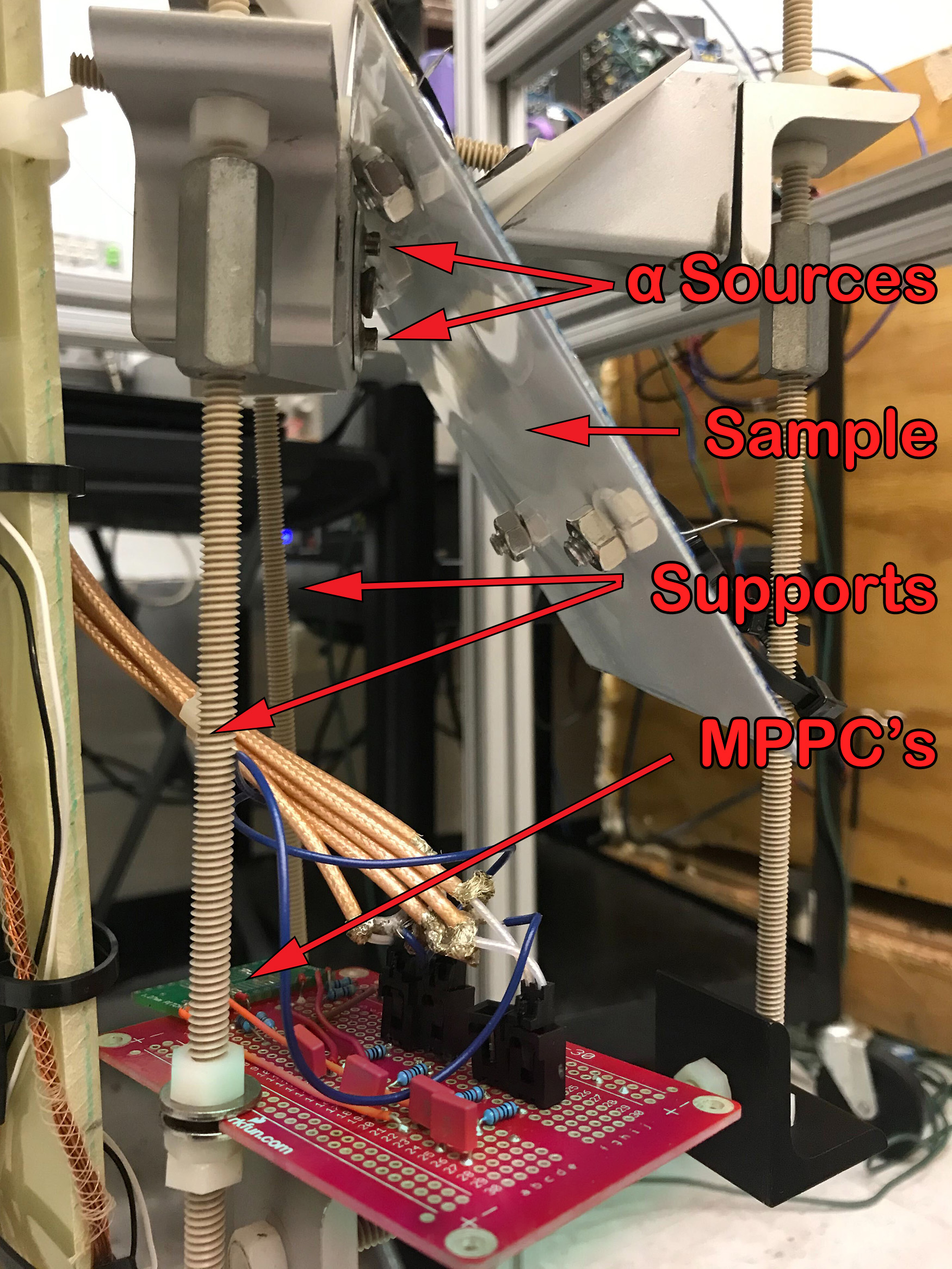} 
\includegraphics[trim = 0.0cm 0.0cm 0.0cm 0.0cm, clip=true, width=0.465\textwidth]{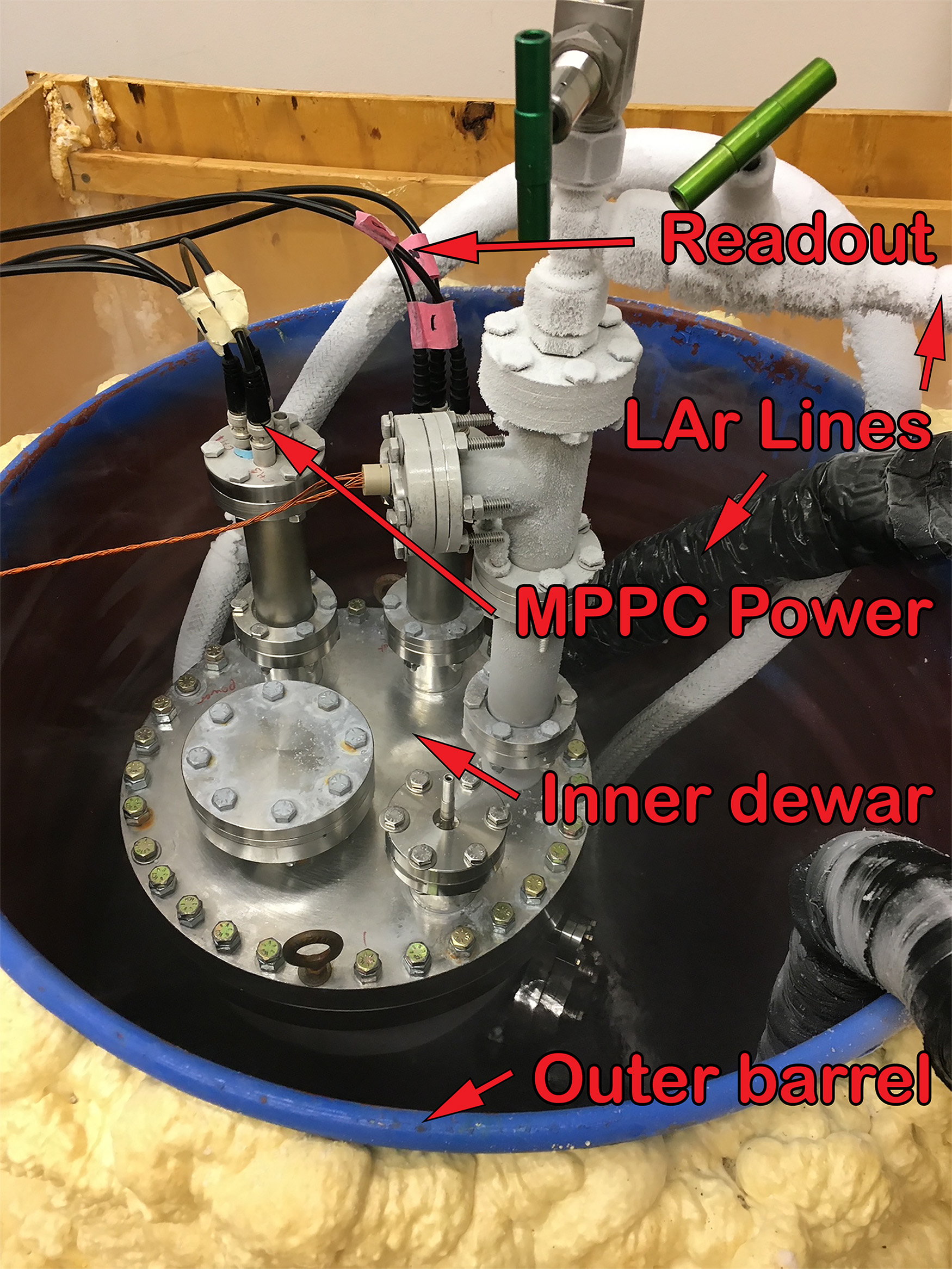} 
    \caption{Left: a PEN-containing sample (top, middle right) as mounted in the testing apparatus, where it faced the MPPC board (bottom green board in back) and adjacent $^{241}$Am sources (top left). Right: the inner cryostat volume, pictured with power and data cables and an open-topped outer barrel.}
    \label{fig:MPPCsAndDewar}
\end{figure}

The experimental setup is visible in~Figure~\ref{fig:MPPCsAndDewar}.  
The vessel hosting the experiment is a light-tight, vacuum-sealable cylindrical inner cryostat roughly 0.07~m$^3$ in volume.  
Feedthroughs in the cryostat top enable delivery of electrical power and LAr into the vessel and argon boil-off and experimental data streams out of the vessel while maintaining good light-tightness and argon purity during data-taking.  
Liquid levels in the cryostat interior are monitored via a series of resistive thermal devices (RTDs) spaced at roughly 12.7~cm (5~inch) intervals.  
The inner cryostat is hosted inside an open-top insulated outer barrel which, when filled with liquid, serves as an insulating thermal bath for the inner cryostat region.  
Liquid supply lines running to both inner and outer volumes enable filling of both regions from a common LAr supply dewar.  

To measure VUV WLS capabilities of test samples in LAr, radioactive sources, light collection elements, and samples were rigidly supported inside of the LAr-filled inner cryostat.  
These elements and their mechanical support system are also visible in Figure~\ref{fig:MPPCsAndDewar}.  
VUV scintillation light is produced using two 0.9 $\mu$Cu Americium-241 ($^{241}$Am) sources mounted on a metal disk in the inner cryostat.  
$^{241}$Am primarily emits $\sim 5.5$~MeV $\alpha$ particles, which ionize and excite argon atoms in the immediate vicinity of the source, causing emission of 128~nm scintillation light.  
To enable greater flexibility in testing the system's operation and response, UV light from a cryostat-external pulsed 250~nm LED circuit could, when desired, be delivered via an optical fiber that terminated a few~cm from the alpha source pointing directly at the test sample.  

Visible light produced as a result of WLS of 128~nm VUV scintillation light within the cryostat is detected by three Hamamatsu S10362-11-050P multi-pixel photon counters (MPPCs) situated at the bottom of the cryostat.  
The wavelength sensitivity and quantum efficiencies of the MPPCs are reported in Ref.~\cite{MPPC}; they are most sensitive in the 350 to 650~nm range, and have no direct sensitivity to 128~nm photons.  
All MPPCs were oriented vertically (light-sensitive elements facing upward) with no direct line-of-sight to the $^{241}$Am sources.  
Source-MPPC distances were roughly 18 cm.  

Each WLS reflective test sample was mounted to a backing plate using four bolts, as shown in Figure~\ref{fig:MPPCsAndDewar}, with a direct line-of-sight to both sources and MPPCs.  
Sample surfaces were placed within a few~cm closest distance to the $^{241}$Am sources at a $\sim$45$^{\circ}$ with respect to the source backing plate and the MPPC sensitive surfaces.  
When present, a sample's test WLS surface (PEN or TPB) was oriented facing the MPPCs and sources, such that VUV scinillation light produced at a source could be absorbed by the WLS material and subsequently re-emitted and reflected in the direction of the MPPCs.  
The mechanical supports for these three primary elements were configured such that test samples could be replaced while preserving the relative distance of both the source and the MPPCs over multiple trials.  
All mechanical supports were connected to a thick G10 rod, which was securely fastened to a contact point on the inner cryostat lid.

\begin{figure}[htb]
    \centering
    \includegraphics[trim = 0.0cm 0.0cm 0.0cm 0.0cm, clip=true, width=0.88\textwidth]{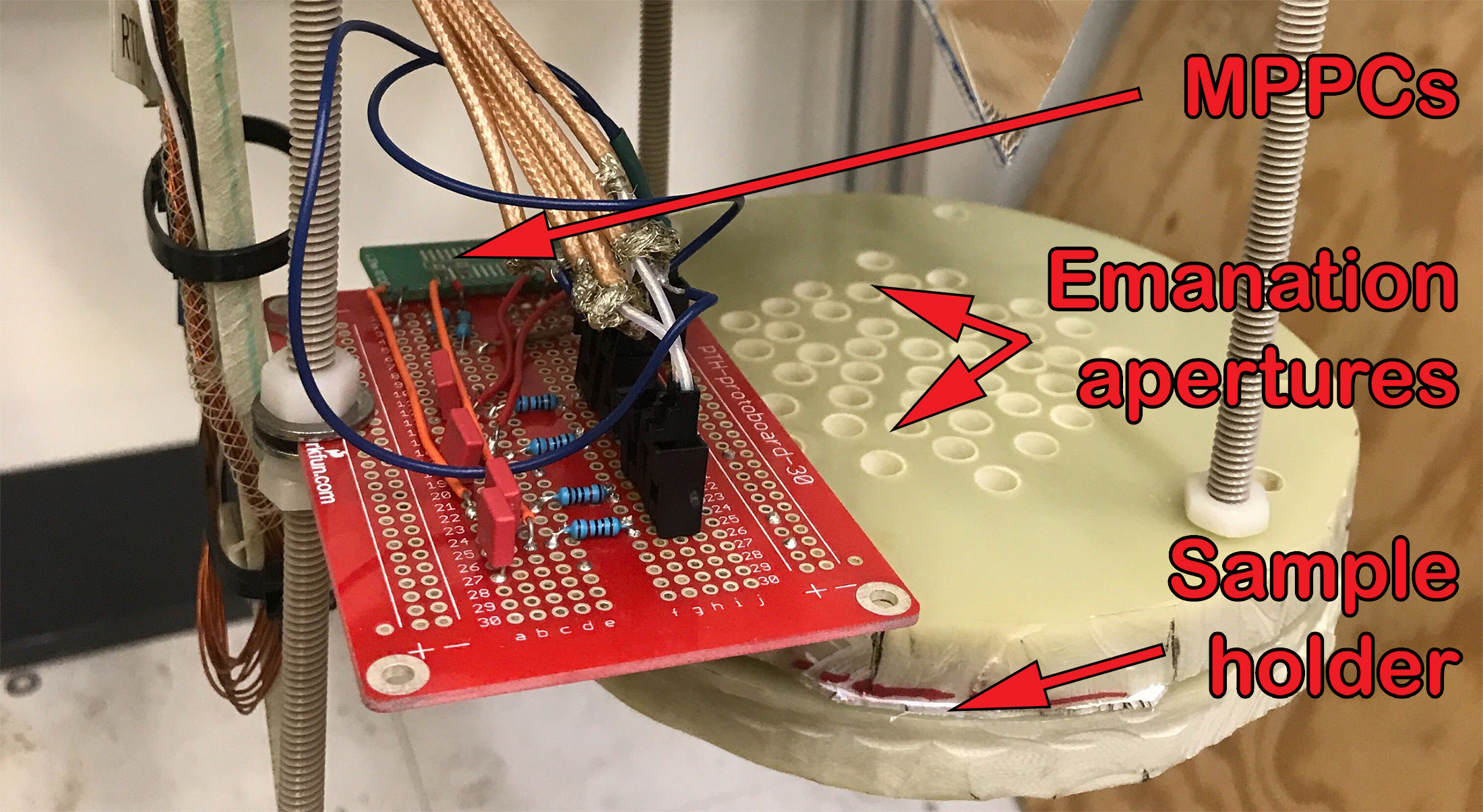}
    \caption{The test setup for WLS material emanation studies. Both MPPCs and $^{241}$Am sources remain in similar locations to Fig.~\ref{fig:MPPCsAndDewar}, while the sample was held below the MPPC board between two hole-filled G10 plates.}
    \label{fig:emanationSetup}
\end{figure}

A subset of data-taking runs were acquired not to compare differences in visible light production by different candidate materials, but instead to check for the emanation of WLS material from the sample into the argon bulk.  
For this series of tests, the sample's orientation was altered as shown in Figure~\ref{fig:emanationSetup}.  
Samples were sandwiched between two thick G10 plates with holes that allowed direct physical contact between sample and bulk argon.  The orientation of the sample below the MPPC board -- as well as a G10 thickness and hole sizes -- did not allow a direct lines of sight between the sample and the MPPCs.  
The orientation achieved for this test was intended to ensure that any visible light collected by the MPPCs was due to VUV WLS by material present in the bulk argon, rather than on the sample itself.  

During data-taking periods, power was provided to the MPPCs via a feedthrough by a TTi PLH120 DC power supply.  
Output signals from each of the three MPPCs were separately carried from the cryostat to three TI-OPA656 pre-amplifiers~\cite{OPA656} mounted close to one of the cryostat feedthroughs.  
These signals were then fed into three channels of a Lecroy Model 612AM NIM-based amplifier~\cite{NimAmp}for further signal enhancement.  
To help remove spurious electronics-based noise, signals were then routed to a band-pass filter designed to filter signals below 1 kHz and above 1 MHz, leaving the signal largely unaffected.  

\begin{figure}[htb]
    \centering
    \includegraphics[scale=0.3]{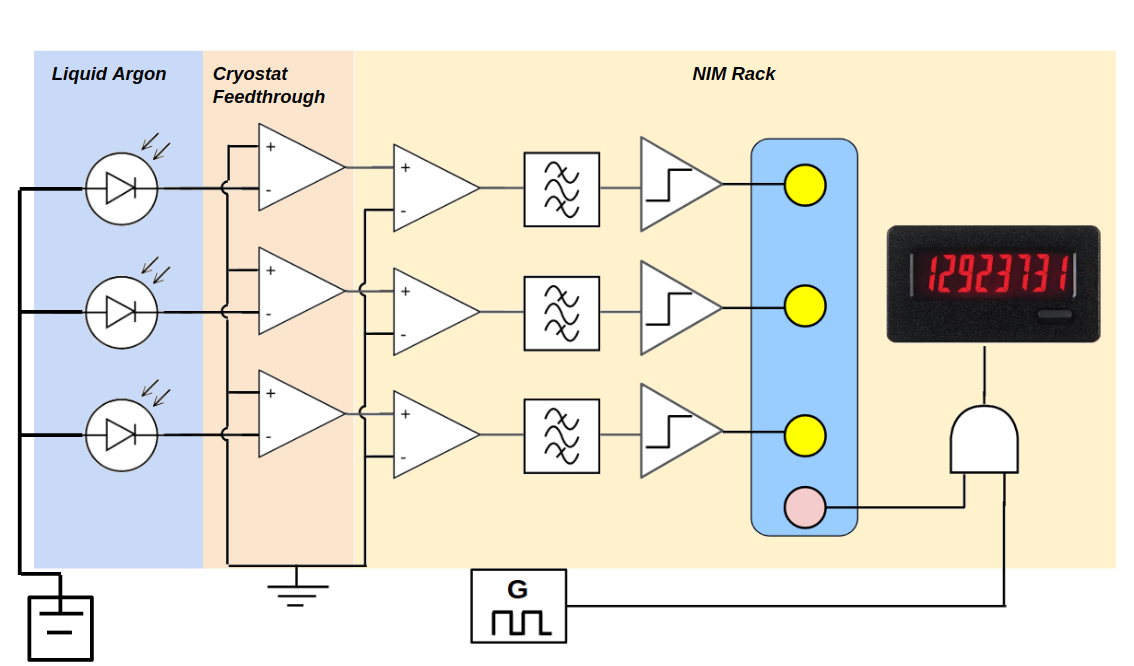}
    \caption{A schematic of the triggering circuit with the MPPCs on the far left and the counts on the right.}
    \label{fig:circuit}
\end{figure}

Filtered pulses were then passed to a NIM-based Lecroy 4608C discriminator~\cite{NimDiscrim}, which triggered on signals above $\sim30$mV for at least 1~ns.  
The discriminator threshold was chosen such that one single photoelectron (SPE) pulse would fire the discriminator while remaining above the residual noise floor.  
After a threshold-crossing, the discriminator transmitted out a 1 volt pulse for 500~ns to a NIM-based Lecroy 365AL coincidence logic unit~\cite{NimLogic}.  
The coincidence logic board took all three discriminator output signals as input and generated an output pulse based on a logic configuration requiring either one, two, or three-fold coincidence.  
When the logic condition was satisfied, a 50~ns output pulse was sent to a NIM-based 10 digit counter~\cite{Counter}.  
The counter was iterated if it also received an input pulse from an external function generator.    
This external function generator was used to accurately define the time window for each trial.  
Once an individual trial was completed, the counter was reset and data taking began again. 
For the data collected and analyzed for this study, a window of 15 seconds was used to count the activity seen by the MPPCs.  
For reference, a diagram of the full readout circuit is shown in Figure~\ref{fig:circuit}.  


\section{Procedures and Systematics} \label{sec:ProceduresSyst}


A data-taking run for one test sample was typically performed over a one or two~day timescale, a period primarily defined by the capacity of two LAr supply dewars typically delivered weekly to the laboratory and the typical boil-off rate from inner and outer LAr volumes.  
In preparation for a data-taking run, the test sample was affixed to the backing plate, and the cryostat lid was carefully lowered and bolted shut from the outside, as seen in Figure~\ref{fig:MPPCsAndDewar}.  
The inner cryostat and argon supply line were then evacuated with a Edwards XDS 15i vacuum pump to $<$10 mT and subsequently filled from the LAr supply dewar, with a pressure relief valve opened to facilitate argon boil-off.  

Argon used for these studies was designated by its supplier as 99.9999\% purity, with impurity specifications of $<$1.0~ppm for O$_2$ and H$_2$O and $<$5~ppm for N$_2$.  
Tests performed on this LAr product using a 10~cm length purity monitor~\cite{ub_det,pm} showed no change in measured electron lifetime before and after active re-circulation and purification, within the monitor's $\mathcal{O}$(1~ms) measurement precision.  
This measurement indicates LAr impurity levels orders of magnitude lower that the quoted specifications, as well as negligible expected levels of impurity-related light quenching or absorption~\cite{warp_o2,warp_n2,n2_abs}.  
Despite the comparatively lower specifications, this high level of demonstrated purity is consistent with the supplier's expectations for this LAr product.  

The inner cryostat and outer barrel were filled simultaneously, with fill level in the inner cryostat defined by the cryostat-internal RTD signatures.  
Liquid levels during filling were not actively equalized; due to its smaller volume, the inner cryostat was the first volume of the two to fill completely.  
After completing filling of the outer barrel, the LAr supply was halted, and the setup was allowed to sit for roughly a half hour prior to data-taking.  
The outer barrel's LAr level was topped up roughly every few hours until the supply dewar for that day's running was exhausted.  
For long-term runs, additional LAr supply dewars were obtained for that week's running, and the outer LAr volume was topped of every few hours for as long as 70 hours of continuous data-taking.  
During both shorter and longer runs, topping off of the inner cryostat LAr volume was not performed.  
Usually only one run was performed for every one week of calendar time.  
Data-taking was performed throughout the 2020 calendar year and in January and February 2021.  

Data-taking for one run began after LAr filling by checking for noise in the system.  
Count rates were observed over a period of 30~s with the MPPCs powered off to verify a count rate of zero, indicating an absence of noise in the readout electronics.  
This was often followed by observations of dark count rates with no source present to estimate this contribution to the count rate, as well as related systematic errors.  
The data-taking procedure for dark count and source measurements was the same: counts in single-hit mode were observed ten times with short breaks in between (15~s measurement period, 15~s for system/electronics re-set), usually followed by the same procedure with the logic board switched to double-coincidence mode and then to triple-coincidence mode.  
This sequence of 10 measurements in each mode, each referred to as a `trial,' was then followed by a longer break of $\sim$1-2 hours. 
For each mode, the mean and standard deviation of each trial's runs were calculated, and the process was then repeated for a total of five trials in each mode.  
Periodically during dark count data-taking, proper operation of the system was verified by  delivering 250~nm LED light to the system and checking for corresponding increases in observed count rates.  

\begin{table}[hptb!]
      \centering
    \begin{tabular}{|c|c|c|}
    \hline
       \textbf{Date of Data Taking}  & \textbf{Mean Counts} & \textbf{Statistical Error} \\
       \hline
       \hline
        Feb 11,2020 & 41.36 & 1.0\\
        \hline
        Feb 13,2020 & 36.5 & 1.9\\ 
        \hline
        Feb 21,2020 & 38.8 & 3.0\\ 
        \hline
        July 13,2020 & 52.98 & 2.6\\ 
        \hline
        July 20,2020 & 58.56 & 2.7\\
        \hline
        July 24,2020 & 80.3 & 2.3\\
        \hline
        Feb 5, 2021 & 52.3 & 2.2 \\
        \hline
        
    \end{tabular}
    \caption{Dark count rates and statistical errors for source-free runs collected at various times during the data-taking period. The statistical error is estimated as the standard deviation of the mean of five dark count measurements, spread over eight hours.}
    \label{tab:DarkCountTable}
\end{table}

Means and statistical errors for different single-hit dark count trials taken at different points during 2020 are shown in Table \ref{tab:DarkCountTable}.  Mean single-hit count rates are relatively stable over that period, with an average rate of 51.4, a standard deviation of the mean of 6.75, and a range of 43.8.  
This full range is conservatively assigned as the uncertainty in the single-hit dark count contributions for all other data-taking runs.  
We treat this uncertainty as uncorrelated between all trials, as well as also considering a similar contribution that is correlated between trials, but uncorrelated between different test sample runs.  
We note that zero dark counts were almost always observed while operating in double-coincidence or triple-coincidence modes.  

When dark count measurements were finished for a particular sample, the system was emptied and opened, $^{241}$Am sources were mounted, and the cryostat was re-sealed and re-filled in the usual manner described above.  
The data-taking procedure described for dark count runs was then repeated with another series of measurements in all three modes over the course of roughly 6 to 8 hours, for a total of five trials for each sample.  

\begin{figure}[htb]
    \centering
    \includegraphics[trim = 0.0cm 1.7cm 0.0cm 1.2cm, clip=true, width=0.6\textwidth]
    {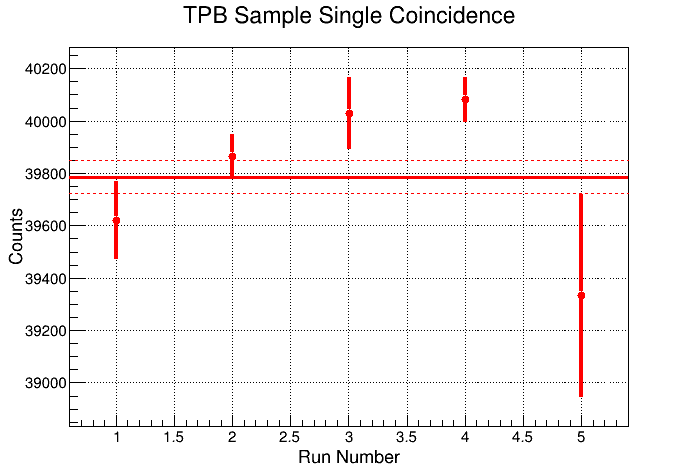}
    \includegraphics[trim = 0.0cm 1.7cm 0.0cm 1.7cm, clip=true, width=0.6\textwidth]
    {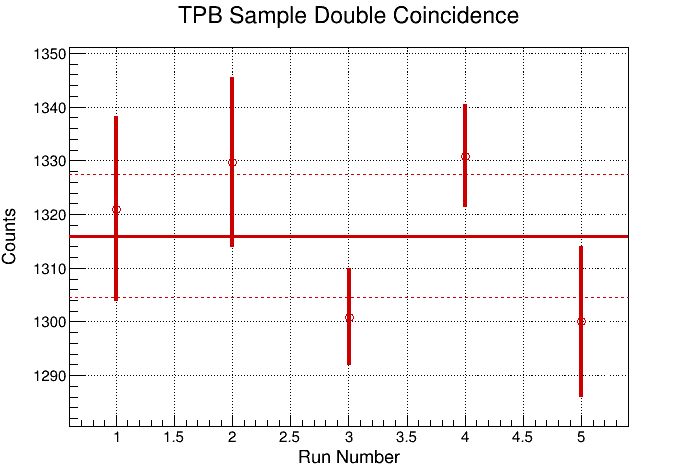}
    \includegraphics[trim = 0.0cm 0.0cm 0.0cm 1.2cm, clip=true, width=0.6\textwidth]
    {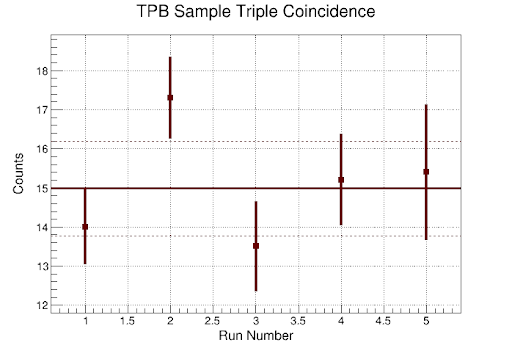}
    \caption{Single-hit (top), double-coincidence (middle) and triple-coincidence (top) mode count rates for the TPB-coated reflector sample.  Solid horizontal lines indicate the mean count rate for each mode, while dotted lines indicate per-trial uncertainties expected from Poisson statistical fluctuations.  All modes generally exhibit stable behavior across trials; detailed quantitative descriptions of system stability are provided in the text.}
    \label{fig:stability}
\end{figure}

To illustrate the appearance of the data and systematic uncertainties contributing to these measurements, average counts per measurement for each mode from each of the five sequential trials of the TPB-containing reflector reference sample are shown in Figure~\ref{fig:stability}.  
Error bars on each point represent the standard deviation of the mean for each trial's measurements.  
The mean in each mode is also shown, along with the expected Poissonian statistical fluctuation given this mean.  
It is reassuring to note that for all data-taking modes, the standard deviation of the mean for each trial, as well as the half-range of all trials, appear to be generally consistent with that expected from Poisson statistical fluctuations and dark count rate fluctuations.  
However, given that the single-hit trials' half-range of roughly 370 counts is roughly five times larger than the expected 77 counts, we conservatively assign this half-range as a 1\% relative uncertainty contribution for each sample run.  
We note that in the analysis that follows, we forego further consideration of double- and triple-coincidence mode data due to their large associated statistical uncertainties compared to single-hit mode.

Qualitatively, the statistical spread between data points in Fig~\ref{fig:stability} for all three modes appears to be larger than any systematic time-dependent drift upward or downward in count rates, indicating that, for shorter measurement periods, coherently time-varying systematics, such as decreasing LAr purity, supply HV drifts, or readout electronics drifts, have only a minor impact on measured light collection.  
Long-term drift was investigated in more detail by observing changes in count rates over a 70-hour data-taking run with the emanation testing geometry described in the previous section; a downward drift of 3 counts per hour was observed when testing the bare ESR reflector sample.  
Further investigation revealed readout system instability as the cause of this drift: by feeding function generator pulses with varying frequency directly into the system, a (0.2$\pm$0.1)\% per hour downward drift was measured in both low-count and high-count scenarios.  
While crucial to correct for and consider in long-term emanation studies, this drift and its uncertainty can be safely ignored for the  shorter-duration relative light collection measurements.  

Run-to-run count rate fluctuations from possible batch-to-batch variations in  liquid purity (and subsequent impurity light quenching), as well as those from geometric or other unconsidered system variations, were probed directly by performing repeated measurements of test samples at different points during data-taking.  
For a Teonex Q53 including test sample, mean single-hit rates during separate runs in mid January 2021 and mid February 2021 were identical within 1.8\%.  
Twin LED light measurements of a bare ESR reflector sample taken in July 2020 and February 2021 provided mean single-hit count rates similar to within 0.8\%.  
A 2\% additional systematic uncertainty is applied to each run to account for  these demonstrated limitations in repetition of light collection results.  
As light collection results for lower-performing PEN01 and PEN02 samples  were not verified with repeat measurements, we conservatively assign an additional 40\% systematic uncertainty for these samples to account for the possible presence of scintillation quenching from anomalously high LAr impurity.  
This 40\% value is obtained by matching the LAr product O$_2$ and N$_2$ impurity specifications given above to quenching factors reported in Refs.~\cite{warp_o2,warp_n2}.  


To briefly summarize, we find that, for all samples, systematic uncertainties dominate statistical errors in measured light collection.  
For measurements of relative light collection between different PEN and reference samples, the dominant systematic uncertainty is attributed to the demonstrated run-to-run repeatability of the system after repeated resets (emptying, opening, closing, and re-filling).  
For extended-run WLS emanation studies, readout electronics drift systematics are dominant.  

\section{Results and Discussion} \label{sec:results}

The results of these studies fall into two categories: measurements of relative wavelength shifting efficiencies of the tested films, and measurements of the long-term stability of samples in LAr.  
The relative efficiency measurement presented in Section~\ref{sec:RelEffic} shows that PEN-including samples enabled visible light collection that was at best 34\% of that from a TPB-including reference, with widely varying performance between different PEN-including samples. 
The stability measurement discussed in Section~\ref{sec:stability} shows that, unlike for TPB, direct PEN-argon contact does not appear to result in collection of visible light from the argon bulk. 


\subsection{Relative Wavelength Shifting Efficiency Results}\label{sec:RelEffic}


Given the stability of geometry and system response demonstrated in the previous section, a comparison of the average counts between the TPB-containing and PEN-containing samples should reflect the relative difference in visible light present in the system in the vicinity of the MPPCs.  
Variation in detected visible light levels between samples could arise from two primary causes described in previous sections: 
\begin{itemize}
\item{Variations in efficiency of shifting 128~nm LAr scintillation light to visible wavelengths detectable by the MPPCs.}
\item{Differences in sample optical properties, particularly bulk attenuation/scattering, surface quality, and refractive indices and index interfaces}
\end{itemize}
Differences in both efficiency and optical properties may arise from the chemical identity of the material (PEN versus TPB), as well as from variations in the polymer structure of the material (biaxially oriented versus amorphous; variations in polymer chain length or level of degradation of polymer structure).  

\begin{figure}[htb]
    \centering
    \includegraphics[trim = 0.0cm 0.0cm 0.0cm 0.0cm, clip=true, width=0.72\textwidth]
    {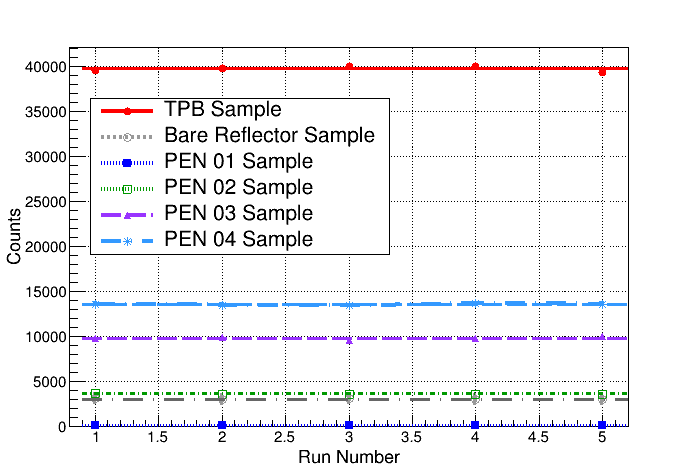}
    \caption{Alpha source data, showing the relative performance of various samples.}
    \label{fig:alphaData}
\end{figure}

Average single-hit counts registered by the MPPCs and electronics per 15~s measurement for each of the tested samples are summarized in Table~\ref{tab:alphaData}, with results from all trials and samples pictured in Figure~\ref{fig:alphaData}.  
Dark count rates, described in the previous section, have been subtracted from all quoted count values.  
Count rates varied widely between samples, ranging from only marginally above dark count rates (PEN01 sample) to roughly 40,000 counts (TPB-coated reflector sample).  
Similar relative count rate differences between samples are also present in double-coincidence data-taking mode, albeit with substantially higher statistical uncertainties; for this reason, we only report single-hit rates for the remainder of this section.  
We note that our observation of non-negligible 128~nm wavelength shifting efficiency from bare ESR reflector is not unexpected, given its partial PEN composition; similar observations were also reported in Ref.~\cite{TPB_Gerda}.  

\begin{table}[]
    \centering
    \begin{tabular}{|c|c|c|c|c|}
    \multicolumn{5}{c}{Single Coincidence: Alpha Source}\\
    \hline
       \textbf{Data Sample}  & \textbf{Mean Counts} & \textbf{Stat. Error} & \textbf{Syst. Error}  & \textbf{Sample/TPB Ratio} \\
       \hline
       \hline
        PEN01 & 169 & 2 & 81 & 0.4\% $\pm$ 0.2\% \\
        \hline
        PEN02 & 3627 & 9 & 1450 & 9.1\% $\pm$ 3.6\%\\ 
        \hline
        PEN03 & 9822 & 14 & 219 & 24.7\% $\pm$ 0.8\%\\ 
        \hline
        PEN04 & 13544 & 16 & 301 & 34.0\% $\pm$ 1.1\%\\ 
        \hline
        TPB & 39786 & 28 & 876 & 100.0\%\\
        \hline
        Bare & 3047 & 8 & 83 & 7.7\% $\pm$ 0.3\%\\
        \hline
    \end{tabular}
    \caption{Single-hit count data for all tested samples in the presence of the ${241}$Am sources.  Absolute counts are provided, in addition to the light collection rate relative to the TPB-coated reference sample.}
    \label{tab:alphaData}
\end{table}

Of the tested PEN-including samples, PEN04, which contains Teonex Q53 supplied by Goodfellow, shows the highest light collection in response to 128~nm scintillation light, with 13544 average single-hit counts.  
This rate represents a visible light level (34.0 $\pm$ 1.1)\%  as high as that exhibited by the TPB-containing sample.  
This level of WLS efficiency and visible light propagation relative to TPB is comparable to preliminary measurements using PEN-coated light collection elements in the protoDUNE-DP detector~\cite{Neutrino2020PENPoster}.  
It is also similar in magnitude to that projected in Ref.~\cite{Kuzniak2019} based on higher-wavelength vacuum UV spectrometry measurements extrapolated to 128~nm using relative low-wavelength measurements from previous literature~\cite{PEN_Old1}.  
With these comparisons made, it should be stressed that the measurements presented in this paper represent, to date, the most direct and geometry-independent tests of PEN performance relative to TPB.  

\begin{figure}[htb]
    \centering
    \includegraphics[trim = 3.5cm 8.0cm 15.0cm 8.0cm, clip=true, width=0.90\textwidth]
    {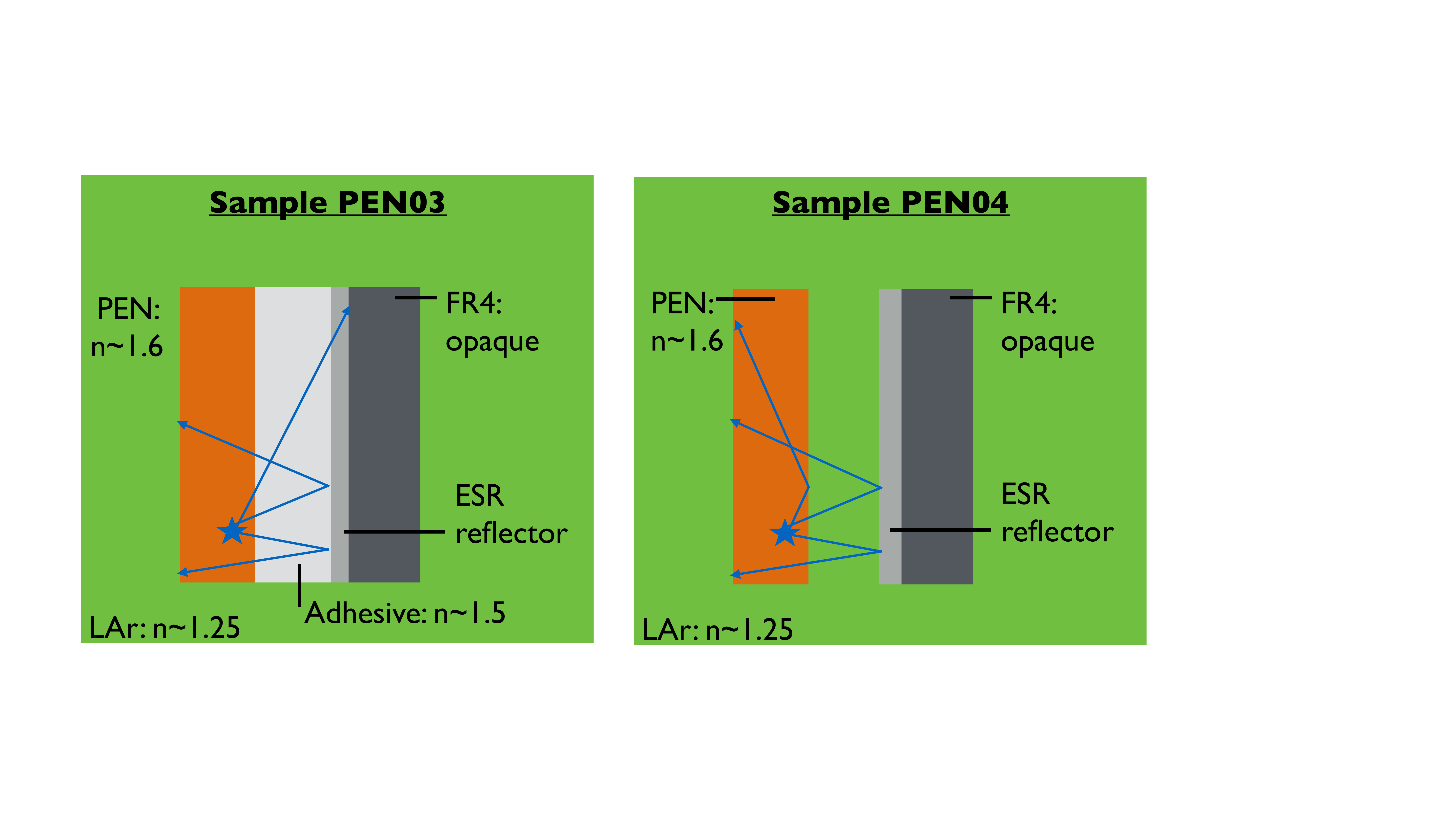}
    \caption{Illustration of difference in optical propagation between PEN03 and PEN04 PEN-containing samples.  Labels indicate the identity and optical properties of the various sample materials, while blue arrows illustrate representative paths of wavelength-shifted light produced in each sample.  In PEN03, light arriving at the back PEN material interface at high incident angle will propagate through the adhesive and ESR reflector, while in PEN04, this light is totally internally reflected back into the PEN sample.}
    \label{fig:TIR}
\end{figure}

Large variations in count rates between samples are also visible in Table~\ref{tab:alphaData} and Figure~\ref{fig:alphaData} between the different PEN-including test samples.  
Average single-hit count rates vary from as low as 170 to as high as 13,544.  
The two best-performing PEN-including samples, PEN03 and PEN04, both contained Teonex Q53.  
Despite the similarity in PEN grade, PEN03 performed (24.7 $\pm$ 0.8)\% as well as TPB, or about 9\% worse than PEN04.  
It is suspected that this difference in performance arises not from relative differences in WLS efficiency of the PEN polymer itself, but rather in propagation of produced visible light to the MPPCs.  
For PEN03, PEN with a refractive index of n$\sim$1.6 was adhered to the ESR reflector via acrylic-based adhesive (n$\sim$1.5), allowing much of the wavelength-shifted visible light produced in the PEN to propagate to the ESR surface.  
As ESR optically coupled to high-index materials has been shown to exhibit high transparency to light at high incident angles~\cite{optical_grid,stereo}, it is likely that a substantial portion of wavelength-shifted light is transmitted through the ESR and absorbed by the sample's FR4 backing.  
For PEN04, the absence of adhesive between ESR and PEN enables lower-index (n=1.25) LAr to separate the materials, resulting in total internal reflection of light above the PEN-LAr interface's critial angle of roughly 50$^{\circ}$.  
Thus, PEN04 is likely to exhibit less loss of light via transmission through the ESR, and a higher rate of collection at the MPPCs.  
An illustration of the proposed differences in optical propagation between PEN03 and PEN04 samples is provided in Figure~\ref{fig:TIR}.  

This hypothesis regarding the difference in performance between PEN03 and PEN04 was tested by performing measurements of one sample using the same source PEN as PEN03 (Teonex Q53 from Millipore Sigma), but reproducing the optical interfaces of PEN04 (no adhesive between PEN and ESR).  
A mean single-hit count rate of 13061$\pm$291 was observed for this sample, substantially higher than that observed from the adhered PEN03 sample.  
This level of light collection is (96.4$\pm$3.0)\% of that exhibited by PEN04, largely validating the hypothesis.  
Thus, it appears that the presence of comparatively low-index LAr behind PEN or other high-index WLS materials can substantially enhance light collection from WLS reflectors.  
More generally, these measurements also indicate that future R\&D studies further exploring and optimizing the optics of PEN-including and other wavelength-shifting reflectors would clearly be worthwhile.  
In particular, future studies should determine if similar enhancements are achieved for other low-index materials, other reflectors, and other WLS candidates.  

The similarity of PEN03 and PEN04 light collection after equalizing their optics also suggests that PEN aging and storage conditions played a sub-dominant role in the relative performance of samples in this study.  
In this case, when all aspects of a tested sample were held constant, excepting the age, storage conditions, and supplier of Teonex Q53 PEN, light collection results were found to be similar within (3.6$\pm$3.0)\%.  
Of course, further systematic studies of light collection variations with aging, humidity, and UV exposure should be performed to more precisely quantify the impact of these variables on PEN thermoplastics.  

The performance of the remaining two PEN-including samples, PEN01 and PEN02, was substantially weaker than the other PEN-including samples described above.  
PEN01, which contained Teonex Q65FA grade PEN, exhibited extremely poor performance, with a light collection only 0.4\% that of the TPB-including reflector.  
This poor performance may be attributable to the proprietary coating applied to Teonex Q65FA which may hinder its VUV WLS performance~\cite{Q65FA_film_application}.  
However, we cannot rule out the culpability of other proprietary details related to Q65FA's production.  
Whatever the cause, it is clear that despite its comparatively superior optical properties, Teonex Q65FA should not be used in LAr-based wavelength-shifting applications.  

Meanwhile, PEN02, which included the only thick amorphous PEN sample, exhibited count rates (9.1 $\pm$ 3.6)\% that of TPB.  
Regarding the lower WLS capability of PEN02, there are a few possibilities which must be considered.  
With respect to sample thickness: the attenuation length of PEN02 has been reported to be $\sim$1 cm at 425 nm~\cite{ORNL_PEN_2019}.  
While this appears unlikely to fully describe the poorer relative performance of PEN02, this attenuation should result in a $\sim20\%$ loss of light compared to samples PEN03 and PEN04.  
PEN02 is also unique in its molecular orientation relative to PEN03 \& PEN04.  
PEN03 and PEN04 samples are biaxially orientated, meaning the dicarboxylate naphthalene moieties are aligned within stacked planes. 
In contrast, PEN02 is amorphous, with a random orientation.  
It has been noted in a similar polymer, poly(ethylene terephthalate), that fluorescence intensity increases with increasing molecular orientation (both uniaxially and biaxially orientated)~\cite{PET_photophysical}; this is proposed to be due to an increased mobility of monomer excitation in the planar geometry. 

\subsection{Emanation Results}\label{sec:stability}

The stability measurement performed on PEN-including samples was identical in many respects to a previous study carried out on TPB in 2018 and described in Ref.~\cite{Asaadi:2018ixs}.  
In the previous study, TPB samples hidden from optical view and from 128~nm scintillation light but exposed to LAr were shown to produce increasing MPPC count rates over days-long time periods.  
This observation suggested the transfer of some quantity of TPB from the sample to the LAr bulk, a hypothesis bolstered by subsequent observation via gas chromatography mass spectrometry of TPB present in filter material through which the LAr was circulated~\cite{gcms}.  
In the current section, we repeat the previous test procedure from Ref.~\cite{Asaadi:2018ixs} with the PEN04 sample to determine whether thermoplastic PEN polymer emanates in a manner similar to evaporatively deposited TPB.  

\begin{figure}[htb]
    \centering
    \includegraphics[trim = 0.0cm 0.0cm 0.0cm 0.0cm, clip=true, width=0.80\textwidth]{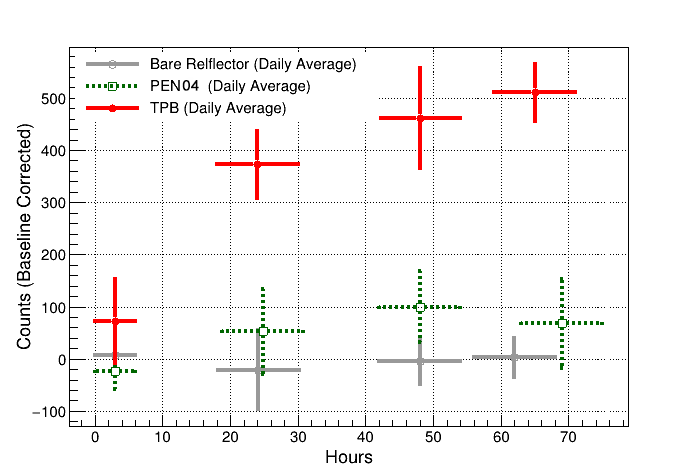} 
    \caption{Baseline-subtracted single-hit count rates versus time for PEN04, TPB, and bare ESR reflector samples in the emanation study geometric arrangement with $^{241}$Am sources deployed. Count rates are corrected for drifts in the baseline over time.  Data error bars represent the statistical spread in measurements, while bands correspond to the uncertainty in the electronics drift correction. } 
    \label{fig:emanationAlphaPlot}
\end{figure}

Average single-hit MPPC count rates recorded at various times during 70-hour $^{241}$Am source run exposures of LAr to various reflector samples are summarized in Figure~\ref{fig:emanationAlphaPlot}.  
Included for testing were the TPB-including and bare ESR samples, as well as the PEN04 sample.    
For all samples, count rates are pictured after subtracting the baseline number of counts observed in the first trial at t=0 hours (which vary between roughly 1300 and 2000 counts), and after correcting for the previously-mentioned (0.2 $\pm$ 0.1)\% per hour downward drift in counts caused by the readout electronics.  
For the TPB-including sample, a detectable increase in counts is observed, as expected: after drift correction, a count rate of 70 at 1.5 hours increases to roughly 500 counts after 70 hours.  
With the ESR reflector sample present, drift-corrected count rates were consistent with zero throughout the full 62~hour run, as expected.  
When testing the PEN-including PEN04 sample, count rates also remained low -- below 100 throughout the 69~hour run.  
Fitting a linear trend to the PEN04 data, we find a best fit slope of (+2.6$\pm$0.26) counts per hour, within 1.3 standard deviations of zero after considering the 2 counts per hour electronics drift uncertainty.  
Data taken concurrently with the pulsed UV LED system similarly showed negligible change in single-hit count rates for bare reflector and PEN sample deployments, along with consistent increases in TPB count rates.  
Thus, it seems clear that thermoplastic PEN films emit far less WLS material than evaporatively deposited TPB; further, we see no statistically significant indication of any PEN emissions whatsoever.   





\section{Summary} \label{sec:discussion}
Liquid argon dark matter and neutrino experiments rely on efficient collection of 128~nm scintillation light to achieve many of their physics goals.  
As these projects scale in size and sensitivity, cost and ease-of-use of WLS coatings and reflecting surfaces will become increasingly important considerations.  
This motivates detailed performance studies of commercially available, easily applicable WLS materials, such as thermoplastic PEN films.  

Using a system with well-demonstrated geometric and response stability, we have measured rates of collection of visible photons produced by wavelength-shifting reflector samples in response to LAr scintillation light.  
Variations in collection rates between test reflector samples arise from differences in wavelength shifting efficiency as well as differences in optical qualities.  
Visible light collection from PEN-including reflector samples is found to be as much as 34\% of that achieved from a TPB-including reference sample.  
This level of performance indicates that PEN may be a viable material for consideration in large-area WLS reflectors or surfaces in future massive LAr particle detectors.  
Additionally, unlike evaporatively coated TPB surfaces, PEN thermoplastic surfaces do not appear to produce substantial emanation of WLS material into the LAr bulk.  

Observed variability in light collection between different PEN-including samples emphasizes  some important characteristics of WLS solutions not previously discussed in the literature.  
First, the level of visible light emitted from a PEN reflector sample appears to be dependent on the polymer structure of the PEN: biaxially oriented PEN films enabled light collection rates 20-40\%  that of TPB, while PEN samples with a more amorphous polymer structure generated $<$10\% light collection relative to TPB.  
Second, light emission from PEN reflector samples is also dependent on the optical interface between the PEN and its backing ESR reflector material: a PEN sample directly adhered to its backing produced a light collection rate  25\% of that exhibited by TPB, while the same PEN grade not directly adhered produced a 34\% rate relative to TPB.  
It is likely that this increase is generated by the presence of low-index LAr between  wavelength-shifter and ESR, which generates highly efficient total internal reflection of WLS light back in the direction of the light collection elements.  
This latter observation suggests that effective wavelength-shifting surfaces may be generated not only from PEN optically coupled to an efficient reflector, but also from PEN optically coupled to a low-index backing material, such as teflon, or optically uncoupled from any backing material.  

\section*{Acknowledgments}
We thank Marcin Kuźniak for discussions and communications about the topic. We also thank José Alfonso Soto Otón for supplying a sample, and for his comments and suggestions. This work was supported by grants from the U.S. Department of Energy, Office of Science, Office of High Energy Physics under Award Numbers DE-SC0008347 and DE-SC0011686, as well as from the Science and Technology Facilities Council (STFC), part of the United Kingdom Research and Innovation; and from the Royal Society UK awards: RGF\textbackslash EA\textbackslash 180209 and UF140089.
We also acknowledge support from UT-Arlington and Illinois Institute of Technology's College of Letters and Science.  

\newpage
\bibliographystyle{JHEP}
\bibliography{bibliography}

\end{document}